\documentclass[twocolumn,nofootinbib]{revtex4-1}
\usepackage{graphicx}
\usepackage{amsmath,amssymb}%,array,dcolumn,subfigure,rotating}

\newcommand{\kbt}{k_{\mathrm{B}}T}

\begin{document}
%%%%%%%%%%%%%%%%

%%%%%%%%%%%%%%%%%%%%%%%%%%%%%%%%%%%%%%%%%%%%%%%%%%
\title{Ion-Specific Hydration Effects: \\
Extending the Poisson-Boltzmann Theory}

\author{Dan Ben-Yaakov, David Andelman}
\email{andelman@post.tau.ac.il}
\affiliation{Raymond and Beverly Sackler School of Physics and Astronomy\\ Tel Aviv
University, Ramat Aviv, Tel Aviv 69978, Israel}
\author{Rudi Podgornik}
\affiliation{Department of Theoretical Physics\\ J. Stefan
Institute, Department of Physics\\ Faculty of Mathematics and Physics and
Institute of Biophysics\\ Medical Faculty, University of Ljubljana\\ 1000 Ljubljana, Slovenia}
\author{Daniel Harries}
\affiliation{Institute of Chemistry and The Fritz Haber Research Center,\\ The Hebrew
University, Jerusalem 91904, Israel}

%%%%%%%%%%%%%%%%%%%%%%%%%%%%%%%%%%%%%%%%%%%%%%%%%%

\date{March 22, 2011 --- submitted to curr. op.}
\begin{abstract}
In aqueous solutions, dissolved ions interact strongly with the surrounding water, thereby modifying the solution properties in an ion-specific manner. These ion-hydration interactions can be accounted for theoretically on a mean-field level by including phenomenological terms in the free energy that correspond to the most dominant ion-specific interactions. Minimizing this free energy leads to modified Poisson-Boltzmann equations with appropriate boundary conditions. Here, we review how this strategy has been used to predict some of the ways ion-specific effects can modify the forces acting within and between charged interfaces immersed in salt solutions.

\end{abstract}
\maketitle

%%%%%%%%%%%%%%%%%%%%%%%%%%%%%%%
\section{Introduction}
%%%%%%%%%%%%%%%%%%%%%%%%%%%%%%%

Resolving the interactions of ions in aqueous solution can be a daunting task due to their complex inter-dependency. Consequently, the theory of electrolytes has been an active field of research for the last century and continues to be so even today. In the early 20$^{th}$ century pioneering achievements  by Gouy, Chapman, Debye, H\"uckel and Langmuir  resulted in the so-called  Poisson-Boltzmann (PB) theory. The starting point of this mean-field theory was to consider point-like ions immersed in a solvent that was modeled as a continuum dielectric. Neglecting charge correlations and fluctuations, the electrostatic energy and the ionic entropy led to impressive predictions on the osmotic pressure, ion activities, and ion profiles close to charged interfaces and electrodes.  These findings still form the basis of much of our current understanding of electrolyte solutions~\cite{RobinsonStokes}.

Even earlier, however, in the late 1800s Hofmeister (among others) encountered ion-specific phenomena when he followed the effects of various salts on macromolecular interactions in solutions~\cite{Hofmeister1}. Specifically, Hofmeister showed that certain monovalent ions (such as fluoride and chloride) are more effective at precipitating proteins (``salting out") than others, such as bromide and iodide. This ion-specific capacity is often described in terms of the ``Hofmeister series" and is correlated with other properties of aqueous ionic solutions, such as their surface tension  and solution viscosity. To explain these observations, one must consider additional ion-ion and ion-solvent interaction, which are not directly and exclusively related to the ion charge. To quote Hofmeister himself ~\cite{Hofmeister2}: ``It also can be expected that the precipitating capacity of salts is parallel to other physical and chemical properties, if [...] these properties are dependent on the water adsorbing capacity of the salts".

In the last decades, the number of observations that found similar ion-specific interactions has steadily grown~\cite{Collins,Conway}. These effects are found to influence the interactions of surfactant micelles, lipid-bilayer membranes, proteins, DNA molecules, and more. As an example we mention charged and net-neutral lipid membranes that can swell or remain unswollen depending on the type of counterions~\cite{petrache1}. This observation can be related to the ``stickiness" of larger and more polarizable ions at the vicinity of membranes. Ion-specific interactions result in a net exclusion or inclusion of ions at interfaces, which then can drive macromolecular interfaces, such as lipid membranes, to bind or separate. An extensive overview is beyond the scope of this review, but the interested reader should consult the recent surveys~\cite{Kunz,Kunz2010}.

Beyond their polarizing effect on water molecules due to their charge, ions are themselves polarizable, and therefore interact with water and interfaces through van der Waals (dispersion) forces. These forces are ion specific, and depend on the ion polarizability. In addition, ions are far from being simple point-like particles, and also show steric interactions that depend on ion size, as quantified by their effective ionic radii. Once we consider more complex ions that are composed of several nuclei, interactions between permanent dipoles should  be examined. Furthermore, these ions can have more complex structure and manifest additional forces, such as hydrophobic interactions between water and nonpolar parts of the ions. Since all these interactions are mediated by the solvating water molecules, we will jointly refer to all these additional interactions under the broad name of {\it hydration forces}, without fully defining their microscopic origin. The focus of this review is on the ways
hydration interactions can be incorporated and used through systematic phenomenological modifications of the PB theory.

\begin{figure*}
  \includegraphics[width=0.8\textwidth]{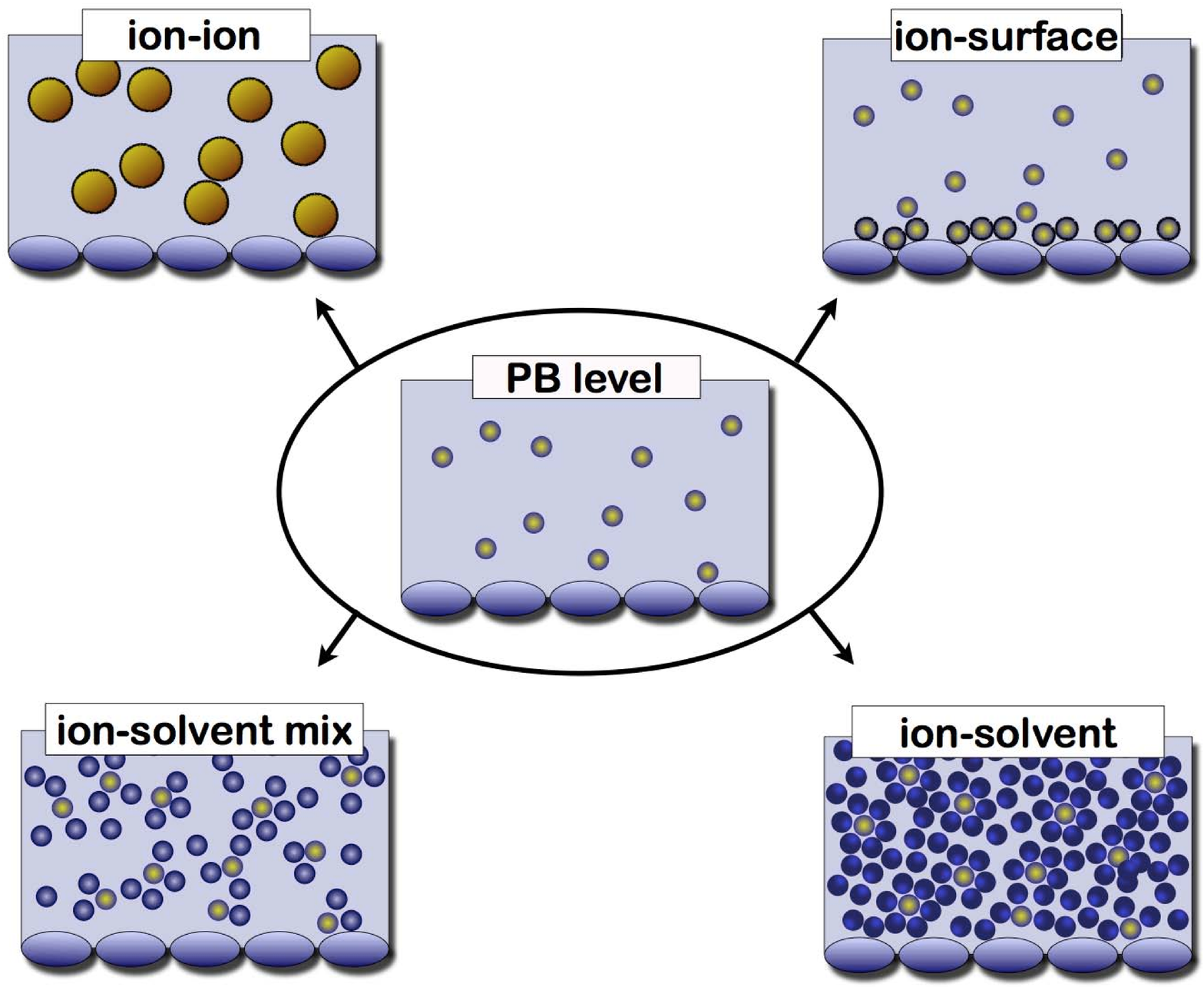}
  \caption{\footnotesize\textsf{Schematic representation of four possible additions to the Poisson-Boltzmann free energy that account for different ion-specific hydration interactions. Additional interactions can involve interactions between the ions themselves (top left),
interactions of ions with surfaces and interfaces that are non-electrostatic (top right), as well as interactions with the solvent that itself could be either single (bottom right) or a multi-component mixture (bottom left).}}\label{fig1}
\end{figure*}

The complexity of hydration interactions makes it almost impossible to account for all their sources {\it ab initio}. The structure of liquid water and hydrogen-bonded network is in itself highly complex to model and predict. Even strategies that rely on molecular dynamics simulations inherently include some level of approximation~\cite{Jungwirth}. As an alternative, we discuss a  phenomenological strategy, where the free energy is our primary tool. The principal idea is that any interaction, including hydration interactions, can be added to the PB free energy, and its effect on the osmotic pressure and ion profiles can be subsequently assessed. Schematic drawing of some of these additional interactions is presented in Figure~\ref{fig1}. In these endeavors we will, however, confine ourselves to short-range hydration interactions that allow for the most straightforward analysis. Similar strategies have been successfully used in other instances where details of specific interactions are not fully known, such as in lipid membranes and liquid crystals.

We start in Sec.~\ref{sec:approach} by presenting the PB free energy of ionic solutions, and the corresponding expressions of the density profile and osmotic pressure. Then, in Secs.~\ref{sec:volume} and~\ref{sec:surface}, we show how additional terms accounting for other non-electrostatic interactions lead to important effects and have consequences to macromolecular interactions.

%%%%%%%%%%%%%%%%%%%%%%%%%%%%%%%%%%%%%%%%%%%%%%%%%%%%%%%%%%%%%%%%%%%%%%%%%%%
\section{General Approach}\label{sec:approach}
%%%%%%%%%%%%%%%%%%%%%%%%%%%%%%%%%%%%%%%%%%%%%%%%%%%%%%%%%%%%
We base our presentation on the PB free energy as the primary starting point~\cite{Andelman},
with its well-known mean-field limitations~\cite{Podgornik1}. For simplicity, let us
focus on a simple model system composed of two charged planar surfaces (of infinite extent) located at  $z=\pm D/2$,
immersed in an electrolyte bath with a dielectric constant $\varepsilon$ that contains  several types of ions.
Each surface carries a charge density of $\sigma$ (elementary charge per unit area). The PB free energy ${\cal F}_{\rm PB}$ of the ionic solution is  composed of  three parts:
\begin{eqnarray}
\label{F_PB1}
{\cal F}_{\rm PB}&=&\int \mathrm{d}^3r \Big[-\frac{\varepsilon\varepsilon_0}{2}(\psi')^2 +   \sum_i e q_i n_i \psi\Big] \nonumber\\
&+& \kbt \int \mathrm{d}^3r  \sum_i \left[n_i \ln(a^3n_{i}) - n_i -\mu_i n_i\right]
\nonumber\\&+& \int_{S} \mathrm{d}^2r \,\sigma \psi\, ,
\end{eqnarray}
where $\varepsilon$ is the dielectric constant, $\varepsilon_0$ is the vacuum permittivity (in SI units), 
and $n_i$ and $q_i$ are, respectively, the number density and the valency of the $i^{th}$ ionic type. The first two terms correspond to the electrostatic energy, while the sum of the third and fourth terms is equal to
$-Ts$, where $T$ is the temperature and $s$ is the ideal
entropy of mixing (per unit volume) of all  mobile ions. Note that $n_i a^3$ is the volume fraction of the $i^{\rm th}$ species, where $a$ is a microscopic length. The chemical potential $\mu_i$ of the $i^{\rm th}$ species, appearing in the fifth term, can be regarded as a Lagrange multiplier ensuring that the ions between the two surfaces are coupled with
an ionic reservoir of a given ionic strength. The last term reflects the surface contribution to the electrostatic energy.

In order to facilitate the generalizations that will follow hereafter,
we rewrite Eq.~\ref{F_PB1} in a more formal fashion:
\begin{eqnarray}
\label{F_PB2}
{\cal F}
&=& \int \mathrm{d}^3r f_V  + \int_S \mathrm{d}^2r f_s \nonumber\\
&=&\int \mathrm{d}^3r \Big( w - Ts -\sum_{i}\mu_i n_i \Big)+ \int_S \mathrm{d}^2r f_s ,
\end{eqnarray}
where $f_V=w-Ts-\sum_{i}\mu_i n_i $ is the generalized bulk free-energy density written as a sum of generalized enthalpy and entropy densities, $w=w(\{ n_i\}, \psi', \psi)$ and $s=s(\{n_i\})$, respectively. In addition, $f_s$ contains all the surface contributions and depends on the surface degrees of freedom such as the surface charge $\sigma$ and surface potential $\psi_s$. The decomposition of the free energy into a volume and surface parts assumes that the surface interactions are of short range.

The Euler-Lagrange (EL) equations of the above free energy can then be decomposed into a volume and a surface contribution. The  bulk equations are
\begin{equation}
\frac{\partial }{\partial z}\left( \frac{\partial f_V}{\partial\psi'}\right)- \frac{\partial f_V}{\partial \psi} = 0\,  \qquad {\rm and} \qquad \frac{\partial f_V}{\partial n_i} = 0\, .
\end{equation}
These EL equations reduce to the common PB equation in the case of  $\cal{F}=\cal{F}_{\rm PB}$ as in Eq.~\ref{F_PB1}. In addition, variation of the surface terms in Eqs.~\ref{F_PB1}-\ref{F_PB2} results in
\begin{equation}
 \left.\frac{\partial f_s}{\partial \psi}\right|_{D/2} =\sigma={\varepsilon\varepsilon_0}\psi_s^\prime .
\label{BC}
\end{equation}
Modifications in any of the free-energy terms in Eq.~\ref{F_PB2} lead to changes in the ionic density profiles and boundary conditions, and represent deviations from the standard PB form.

From the PB free energy, Eq.~\ref{F_PB1}, it follows that the pressure between the confining surfaces is
\begin{equation}
P_{\rm PB}=-\frac{1}{\rm Area}\frac{\delta {\cal F}_{\rm PB}}{\delta D}=-\frac{\varepsilon\varepsilon_0}{2}(\psi^\prime)^2 +\kbt \sum_i n_i\, .
\label{P_PB}
\end{equation}
The  pressure is composed of an attractive electrostatic term stemming from the Maxwell stress tensor~\cite{Dan1}, and a second term that resembles the van 't Hoff form and originates from the ideal entropy of mixing. 

The osmotic pressure is given by the difference between the pressure $P$ at finite separation and its bulk value $P_b$ (infinite separation). In the following we will refer to $P$ as the osmotic pressure, omitting the reference to its bulk value. For systems with symmetric boundary conditions
(equally-charged surfaces), the osmotic pressure is always repulsive~\cite{Neu}. Note that, in general, boundary terms in the free energy {\it do not contribute explicitly} to the  pressure. 

Generalizations of the above electrostatic free energy are represented in Figure 1 
and entail changes in all three of its parts, $w$, $s$ and $f_s$. Modifications in the enthalpy $w$ are related to the dependence of dielectric response function $\varepsilon$ on solvent and solute compositions, as well as with other non-electrostatic interactions. The changes in the ionic entropy $s$ are related with ionic steric interactions. Finally, modifications of the surface free-energy $f_s$ are wrought by the short-range non-electrostatic ion-surface interactions. The approach advocated here provides a convenient way to classify non-electrostatic effects in terms of how they modify the standard mean-field free energy.

The extended bulk free-energy $f_V$ can be written in a rather general but explicit form
\begin{equation}
f_V = -\frac{\varepsilon_0}{2}\varepsilon\left(\left\{ n_{i}\right\} \right)\left(\psi^{\prime}\right)^{2}+
\sum_{i} e q_i n_{i}\psi~+~h\left(\left\{ n_{i}\right\} \right)\,,\label{eq:f}
\end{equation}
where the dielectric response
$\varepsilon\left(\left\{ n_{i}\right\}\right)$ takes a general form and depends on the ionic densities.
The  expression for $f_V$ includes the changes in the electrostatic interaction part (the first term), as well as in the third term, where $h$ combines entropy and all other non-electrostatic interactions.

The corresponding osmotic pressure can then be derived as
\begin{equation}
-P=\frac{\delta {f}_{\rm V}}{\delta D}=\frac{\varepsilon_0}{2}\Big[\varepsilon+\sum_{i}\frac{\partial\varepsilon}{\partial n_{i}}n_{i}\Big]\left(\psi^{\prime}\right)^{2}+h-\sum_{i}n_{i}\frac{\partial h}{\partial n_{i}}.
\label{eq:P}
\end{equation}
Only modifications in $f_V$  enter explicitly into the pressure expression. However, surface modifications lead to changes in the density profiles, which in turn cause deviations in $P$. The function $h\left(\left\{ n_{i}\right\} \right)$  encodes all the non-electrostatic couplings of the ionic densities. Hence, the osmotic pressure depends crucially on the assumed form of $h$ as will be demonstrated in several cases hereafter.

%%%%%%%%%%%%%%%%%%%%%%%%%%%%%%%%%%%%%%%%%%%%%%%%%%%%%%%%%%%%%
\section{Bulk: ion-ion and solvation interactions}\label{sec:volume}
%%%%%%%%%%%%%%%%%%%%%%%%%%%%%%%%%%%%%%%%%%%%%%%%%%%%%%%%%%%%%
In the following section we review several examples where the bulk free-energy, $f_V$,
is modified on the mean-field level, shedding new light on the notion of hydration interactions.

%%%%%%%%%%%%%%%%%%%%%%%%%%%%%%%%%%%%%%%%%%%%%%%%%%%%%%%%%%%%%%%%%%%%%%%%%%%
\subsection{Steric effects of hydration shells}
%%%%%%%%%%%%%%%%%%%%%%%%%%%%%%
Why would one consider steric effects as a part of hydration interactions? In an aqueous solvent, the hardcore radius of ions is set principally not by ionic impenetrability but rather by the much larger and variable size of its hydration shell, composed of highly structured vicinal water~\cite{Conway}. This structured shell excludes ions from the proximity of one another (top-left panel of Figure 1). The full account of this problem can  be addressed by employing liquid state theories, and requires extensive tools such as computer simulations or numerical solution of integral equations~\cite{packing}. However, even with these elaborate tools, the state-of-the-art picture of solvent and solute structure is not well understood.

An alternative approach is to use a more crude, yet simple, mean-field theory that accounts for steric effects within the  Coulomb lattice-gas model. On this level the lattice constant is chosen to be equal to the hydration shell radius, $a$. The model accounts for  density saturation close to charged objects and surfaces, but does not provide any insight into the microscopic packing structure.

\begin{figure}
  \includegraphics[width=85mm]{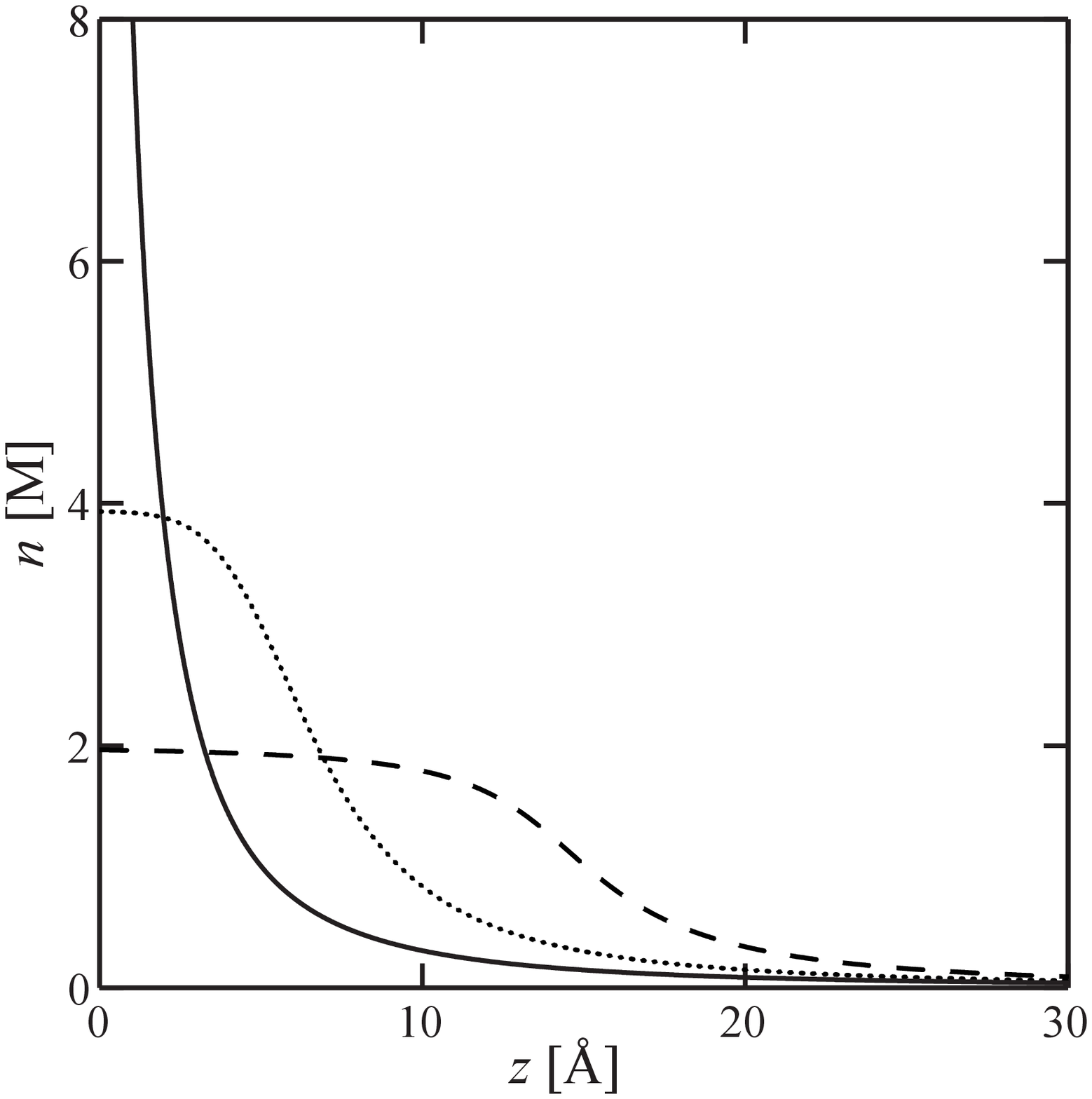}
  \caption{\label{fig2}\textsf{Counterion density $n$ as a function of distance $z$ from a charged interface. Results of three models are presented: the standard PB theory (solid line);  PB theory including ionic hydration shell, Eqs.~\ref{eq:mpb1}-\ref{Psteric}, of size $a=7.5\,$\AA\ (dashed line); PB theory including linearly varying dielectric constant, Eqs.~\ref{eq:lin_eps}-\ref{eq:lin_eps1}, with $\beta=-20\,$M$^{-1}$ (dotted line). The surface charge density  is $-e/50\,$\AA$^{-2}$ in all cases.}
  %{\bf DA: please change green into dashes and red into dots.}
  }
\end{figure}

The lattice-gas model of Coulomb fluids has a venerable history~\cite{Thefrenchguy} and has been apparently reinvented several times in the last century~\cite{Bazant}, starting with  Bikerman in 1942~\cite{Bikerman}. The lattice-gas entropy  has the mean-field form
\begin{eqnarray}
& &TS(\left\{ n_i, \psi \right\}) = - \frac{k_BT}{a^3} \int \mathrm{d}^3r \left[\sum_i  n_i a^3 \ln(a^3n_{i}) \right. + \nonumber\\
& & \left. + \left(1 - \sum_i n_i a^3\right) \ln\left(1 - \sum_i n_i a^3\right)  \right].
\label{eq:mpb1}
\end{eqnarray}
and can be systematically derived via a field-theory approach~\cite{Borukhov}.
For simplicity, it is assumed that all ions have the same hydration shell size $a$ but this condition can be relaxed~\cite{Tresset}. The corresponding  EL equations lead to a modified PB equation whose solutions show a pronounced saturation of the ion density in high potential regions, as can be seen in Figure~\ref{fig2} (dashed line)~\cite{Andelman1}. This limits the highest ionic density in the vicinity of charged surfaces, extending the electrostatic double-layer further into the bulk, as compared to the standard PB profile.

An even older attempt to take into account steric effects at charged surfaces was introduced by Stern in the 1920s~\cite{stern}. In Stern's model, ions are assumed to be fully excluded from a layer of a few Angstroms (the so-called Stern layer) adjacent to the surface, while outside this layer the profile is determined by the standard PB model. The lattice-gas and the Stern models are both related to steric effects, although their ion-density profiles differ at the surface.

The osmotic pressure for a Coulomb lattice-gas between two charged surfaces was derived in Ref.~\cite{Dan1}, having the form
\begin{equation}
P = -\frac{\varepsilon\varepsilon_0}{2}(\psi^\prime)^2 - \frac{\kbt}{a^3} \ln\left(1 - \sum_i n_i a^3\right),
\label{Psteric}
\end{equation}
which can be reduced to the PB form of Eq.~\ref{P_PB} in the limit of a small hydration shell ($a^3 \rightarrow 0$).

We note that steric exclusion, in general, augments the osmotic pressure between charged surfaces  at small separation and provides a paradigm for the ``hydration interaction" that stems from entropy contributions to the free energy.

%%%%%%%%%%%%%%%%%%%%%%%%%%%%%%%%%%%%%%%%%%%%%%%%%%%%%%%%%%%%%%%%%%%%%%%%%%%
\subsection*{Ion-dependent dielectric response}
%%%%%%%%%%%%%%%%%%%%%%%%%%%%%%%%%%%%%%%%%%%%%%%%%%%%%%%%%%%%%%%%%%%%%%%%%%%
The formation of hydration shells around ions is driven by the strong local electrostatic fields~\cite{Conway}. Besides the steric effects (discussed in the previous section) due to the formation of the hydration shell, the dielectric response of the solution is also modified by the presence of ions~\cite{Debye}. Being dipolar, water molecules are strongly attracted to the ions and  aligned themselves along the electrostatic field lines that originate from the ions. Consequently, the orientational degrees of freedom of water molecules are substantially reduced, leading to regions with lower dielectric response compared to pure water (bottom-right panel of Figure 1). This results in a  decrement of the macroscopic dielectric constant as has been observed experimentally already in the late 1940's~\cite{hasted}. Furthermore, when the ions are distributed inhomogeneously (for example, close to charged interfaces), the dielectric response will depend locally on the ionic profile.

In the framework of a phenomenological mean-field free-energy,  this effect~\cite{Dan2} can be accounted for by assuming that the dielectric response depends linearly on the ionic density, $n$:
\begin{equation}
\varepsilon[n(z)]=\varepsilon_w+ \beta n(z)\, ,
\label{eq:lin_eps}
\end{equation}
where $\varepsilon_w$ is the dielectric constant of pure water, and $\beta$ is defined as a phenomenological  coefficient. The linear relation for macroscopic dielectric constant was observed for several monovalent salts and for a fairly wide range of  concentrations, see Ref.~\cite{Dan2} and references therein. Note that only one ionic species (counterion only) is considered here for the sake of simplicity, but this can be easily generalized for any number of ionic species.

The counterions now play a ``dual" electrostatic role. On the one hand, they are attracted to the charged interface due to their charge, whereas, on the other hand, they are repelled due to their lower dielectric response relative to the pure solvent. The interplay between these two contributions leads to a decreased ionic density in the vicinity of the charged interface, when compared to standard PB theory. Specific results can be found in Ref.~\cite{Dan2}.

The above mentioned dielectric response, Eq.~\ref{eq:lin_eps}, leads to an explicit term, $-\beta\varepsilon_0 n(\psi^{\prime})^2/2$, to be inserted in the bulk free-energy, $f_V$ of Eq.~\ref{eq:f}. In the prevalent case for monovalent salts, the linear coefficient is negative,
$\beta<0$, and
the free energy is increased due to the presence of the ions. The corresponding osmotic pressure can be obtained from Eq. \ref{eq:P} and is of the form
\begin{equation}
{P}=-\frac{\varepsilon_0}{2}\left[\varepsilon_w+
2\beta n\right](\psi')^2+ \kbt n\, .
\label{eq:lin_eps1}
\end{equation}
The   $2\beta n\left(\psi^{\prime}\right)^2$ term together with the modified  ionic profile (exclusion effects) results in an increase of the repulsion between like-charged surfaces, as compared to the PB pressure, Eq.~\ref{P_PB}, as can be seen in Figure~\ref{fig2} (dotted line). This may appear as counterintuitive, because exclusion often entails effective attractive interactions between interfaces (so-called depletion interactions). However, it should be realized that the counterions {\it should} remain in the intervening space between equally charged surfaces, and they can only do so at a larger free-energy cost as compared to the standard PB theory.

%%%%%%%%%%%%%%%%%%%%%%%%%%%%%%%%%%%%%%%%%%%%%%%%%%%%%%%%%%%%%%%%%%%%%%%%%%%
\subsection*{Ions in a dipolar solvent}
%%%%%%%%%%%%%%%%%%%%%%%%%%%%%%%%%%%%%%%%%%%%%%%%%%%%%%%%%%%%%%%%%%%%%%%%%%%
An alternative approach to the effect of ion-solvent interaction is to model water molecules as explicit, constant-magnitude point-like dipoles~\cite{DPB,May2010}. This adds another degree of freedom to the free energy of Eq.~\ref{F_PB2}. Since the dipolar degree of freedom can be integrated analytically, one obtains a free-energy ansatz along the lines of Eq.~\ref{F_PB2}, with an additional polarization-energy term. Instead of treating the solvent as a homogeneous medium with a macroscopic dielectric constant, the dielectric constant is calculated to be a function of the local electrostatic field, $\psi'$:
\begin{equation}
\varepsilon_{\rm eff}=\varepsilon_0+ \frac{c_d p_0}{ \psi'}{\cal G}(u)\, ,
\end{equation}
where the bulk concentration of dipoles is $c_d$, $p_0$ is their dipolar moment,  and $u\equiv p_0\psi'/\kbt$ is a dimensionless energy variable accounting for the energy of a dipole $p_0$ in an external electric field $-\psi'$. The function ${\cal G}(u)\equiv  {\cosh u}/{u} - {\sinh u}/ {u^2}$  is related to the first order spherical Bessel function. The osmotic pressure for this model can be computed explicitly and leads to the form
\begin{eqnarray}
P=-\frac{\varepsilon\varepsilon_0}{2} (\psi')^2 + \kbt\sum_i n_i \nonumber\\
-\kbt c_d u {\cal G}(u) + \kbt c_d \frac {\sinh {u}}{ u}\, ,
\end{eqnarray}
where the third term is the dipolar electrostatic contribution, and the fourth term is  the pressure of an ideal gas of dipolar particles with a concentration $c_d \sinh u/u$. The interplay between these two terms as well as the change in the ionic profile leads to a deviation from the PB pressure, making it smaller in the case of two oppositely charged surfaces~\cite{DPB}.

%%%%%%%%%%%%%%%%%%%%%%%%%%%%%%%%%%%%%%%%%%%%%%%%%%%%%%%%%%%
\subsection*{Ions in hydrogen-bonded aqueous solvent}
%%%%%%%%%%%%%%%%%%%%%%%%%%%%%%%%%%%%%%%%%%%%%%%%%%%%%%%%%%
Another ion-specific effect  explicitly considers the hydrogen-bond network of water molecules and its perturbation by the dissolved ions and confining surfaces~\cite{hydref}. These types of considerations lead to finite-ranged hydration interactions and complicated couplings between the chemical structure of the confining surfaces~\cite{hydrationtheory}. Of all the effects considered here these are the most difficult to describe theoretically in a consistent way, since new types of order parameters, in addition to ionic density and the corresponding electrostatic fields, need to be taken into account. Various types of phenomenological theories of finite-range hydration effects are mostly related to the microscopic Onsager-Dupuis model of ice~\cite{marceljaice} and are able to account for some facets of the structural correlations of the aqueous solvent as a consequence of its pronounced hydrogen-bonded structure.

%%%%%%%%%%%%%%%%%%%%%%%%%%%%%%%%%%%%%%%%%%%%%%%%%%%%%%%%%%%%%%%%%%%%%%%%%%%
\subsection*{Ions in binary mixtures}
%%%%%%%%%%%%%%%%%%%%%%%%%%%%%%%%%%%%%%%%%%%%%%%%%%%%%%%%%%%%%%%%%%%%%%%%%%%
Models accounting for ion-solvent interactions can also be generalized for the case of binary solvent mixtures. These are complex solvating environments, because the solution may respond to a solvated ion by varying the local solvent composition (bottom-left panel of Figure 1). For example, a miscible mixture of high dielectric solvent (water) with a lower dielectric one (such as alcohol) can respond to a solvated ion by drawing more water molecules around the ion on the expense of alcohol molecules in order to optimize the dielectric response~\cite{solvation1}. Moreover, additional non-electrostatic interactions may preferentially drive one solvent to make closer contact with the ions. This general tendency is characterized by so-called preferential interactions, which compete with the entropic cost (or more generally, the free-energy cost) of concentrating one solvent around the ion. These interactions can, for example, be quantified by the differences in solvation energies of salts by different pure solvents and can even lead to macroscopic phase-separation of the liquids (``salting out")~\cite{solvation2}. Similar considerations can be applied to charged macromolecular interfaces. For example, recent experiments show that preferential water inclusion (preferential hydration) can lead to subtle, yet measurable changes in the forces acting between macromolecules such as DNA~\cite{Rau2006}.

At the simplest level, the local dielectric response in an A/B binary mixture depends linearly on the local solvent composition $\phi_A$ and $\phi_B$ as 
\begin{equation}
\varepsilon(z) = \phi_A(z) \varepsilon_A + \phi_B(z) \varepsilon_B\, ,
\end{equation}
where due to solvent incompressibility (while neglecting the ionic volume fraction), 
we require $\phi_A+\phi_B=1$.  This form is supported also by experimental evidence~\cite{solvation2}.

The total free energy is composed of three terms: ${\cal F}={\cal F}_{\rm PB}+{\cal F}_{\rm mix}+{\cal F}_{\rm sol}$. The first ${\cal F}_{\rm PB}$ term is the regular PB expression, Eq.~\ref{F_PB1}, with the dielectric constant $\varepsilon$ of the solvent substituted by the local dielectric response, $\varepsilon(z)$. The second term ${\cal F}_{\rm mix}$ is directly derived from regular solution theory, and describes the ideal mixing and enthalpy of the solvent on the mean-field level in terms of the relative solvent composition
$\phi\equiv \phi_B$:
\begin{eqnarray}
{\cal F}_{\rm mix}=\frac{k_BT}{a^3} \int \left[ \phi \ln \phi + (1-\phi) \ln (1-\phi)\right]d^3r \nonumber\\ + \frac{k_BT}{a^3} \int \chi \phi (1-\phi)  \, d^3r\, ,
\end{eqnarray}
where $\chi$ is the dimensionless interaction parameter (rescaled by $k_BT$). The origin of the third term, ${\cal F}_{\rm sol}$, is the preferential (non-electrostatic) ion-solvent interaction, assumed to correspond to a bilinear coupling between the ion densities, $n_i$, and the relative solvent composition $\phi$, so that
\begin{equation}
{\cal F}_{\rm sol}=k_BT \int \sum_i  ~\alpha_i n_i \phi~ d^3r
\end{equation}
where $\alpha_i$ is the preferential interaction parameter of species $i$.

Minimizing the free energy with respect to $\phi$, $n_i$ and $\psi$ yields the corresponding equilibrium (EL) equations \cite{JPCB_us,yoav_sela}.  Solving these equations results in ionic and solvent density profiles close to the charged surfaces. The osmotic pressure can be derived from Eq.~\ref{eq:P} as a function of the inter-plate separation $D$ and the experimentally determined by the bulk salt concentration $n_b$, the bulk solvent composition $\phi_b$, and $\alpha_i$. It has the form
\begin{eqnarray}
P &=& -\frac{\varepsilon_0}{2}\Big[\varepsilon+ \frac{\partial\varepsilon}{\partial \phi}\phi\Big]\left(\psi^{\prime}\right)^{2} + \nonumber\\
&+& \kbt \left( \sum_{i} n_i - \frac{1}{a^3} \ln\left(1 - \phi \right) \right) + \nonumber\\
&+&  \kbt \left( \sum_i  ~\alpha_i n_i \phi- \frac{\chi}{a^3} \phi^2 \right).
\label{Pstericmixed}
\end{eqnarray}
In the absence of preferential solvation, $\alpha_i=0$, and only as long as $\phi_b\ll1$ and $\varepsilon_A>\varepsilon_B$, small changes in osmotic pressure are expected, because density modulations of the solvent are small.
However, considerable changes in the osmotic pressure are expected when preferential interactions  are included \cite{JPCB_us}; the effect is expected to be particularly large for separation between charged interacting surfaces of less than $\sim 10-20$\,\AA.

%%%%%%%%%%%%%%%%%%%%%%%%%%%%%%%%%%%%%%%%%%%%%%%%%%%%%%%%%%%%%%%%%%%%%%%%%%%
\subsection*{Solutions of antagonist ions}
%%%%%%%%%%%%%%%%%%%%%%%%%%%%%%%%%%%%%%%%%%%%%%%%%%%%%%%%%%%%%
Along similar lines as presented in the previous section, an interesting effect emerges by considering the peculiar case where co-ions and counterions bear an {\it antagonist} preferential solvation~\cite{Onuki1,Onuki2}. This effect is of considerable importance to the properties of neutral interfaces where the two ionic species accumulate on different sides of an interface between two immiscible dielectric solvents. For example, we mention the oil/water interface, where the hydrophobic ions accumulate on the oil side of the interface, and  the hydrophilic ones on the aqueous side. Under these conditions the surface tension is found to depend non-monotonously on the ionic bulk density, resulting from a negative contributing term proportional to $\sqrt{n_b}$, and a positive term linear in $n_b$. Note that in order to obtain this behavior even on the mean-field level, one must take into account image charge interactions, which become important near non-charged interfaces separating two different dielectric solutions. Here these interactions were considered in a generalized Onsager-Samaras fashion and are further discussed in Sec.~\ref{sec:surface}. More recently~\cite{Onuki3}, the implication of this effect on the phase diagram of aqueous mixtures was studied. It  was found that due to this antagonist ion effect, hydrophilic ions tend to enhance phase separation between the two immiscible solvents.

%%%%%%%%%%%%%%%%%%%%%%%%%%%%%%%%%%%%%%%%%%%%
\section{Direct ions-surface interactions}
%%%%%%%%%%%%%%%%%%%%%%%%%%%%%%%%%%%%%%%%%%%%
So far, we have discussed the contributions to the bulk energy of solvated ions, $f_V$. Similarly, specific interactions can also occur at interfaces when the adsorbed ions interact favorably with the surface constituents (top-right panel of Figure 1). The main difference is in the lower dimensionality of the adsorbing surface. Below we review some examples of short-range surface-specific interactions that can be described via a modified surface free-energy, $f_s$, (second term in Eq.~\ref{F_PB2}), and examine the consequences of these modifications.

%%%%%%%%%%%%%%%%%%%%%%%%%%%%%%%%%%%%%%%%%%%%%%%%%%%%%%%%%%%%%%%%%%%%%%%%%%%
\subsection*{Charge regulation}
%%%%%%%%%%%%%%%%%%%%%%%%%%%%%%%%%%%%%%%%%%%%%%%%%%%%%%%%%%%%%%%%%%%%%%%%%%%
In their pioneering model of charge regulation, Ninham and Parsegian~\cite{chargereg} introduced the concept of a variable surface charge density that self-adjusts according to other system properties. The charged surface is not described by a constant charge density. Instead, the amount of charge is regulated according to association and dissociation of surface ionic groups. Considering such processes at thermal equilibrium yields a charge regulated boundary condition. The main addition in the Ninham-Parsegian model with respect to the regular PB model is the novel boundary condition, which is determined self-consistently and, in turn, fixes the amount of dissociated charge groups on the surface as a function of, {\it e.g.,} the separation between the surfaces. This model proved to be very useful and was later applied in a plethora of other contexts~\cite{chargereg2}.

Though originally the charge regulation was derived by requiring that the surface charging is at equilibrium with the bulk through a specified equilibrium constant for ion-surface binding~\cite{chargereg}, it can be introduced equivalently by using a free-energy approach. Consider a surface composed of associated (neutral)
and dissociated (charged) groups modeled as a two-component mixture. The free energy, $f_s(\eta, \psi_s) =
\sigma \psi_s + f_{\rm ent}$, is a sum of an electrostatic term and the surface entropy of mixing
\begin{equation}
f_{\rm ent}=\frac{k_BT}{a^2} \left[ \eta \, \ln\eta \right. + \left.
\left(1 -  \eta \right) \ln\left(1 -  \eta\right)  \right],
\end{equation}
where $\eta$ is the area fraction of the dissociated (charged) groups at the surface, and $\sigma = e \eta/a^2$ is the surface charge density. The boundary condition, Eq.~\ref{BC}, is obtained from a variation of the free energy with respect to $\psi_s$, while minimization with respect to $\eta$ corresponds to the Langmuir adsorption isotherm~\cite{Daniel1}. Combining these two conditions leads to the Ninham-Parsegian charge regulation condition~\cite{chargereg}:
\begin{equation}
\frac{\eta}{1 - \eta} = \exp\left(-\frac{ e \psi_s}{\kbt}\right)\, .
\end{equation}

Since the boundary terms do not enter the pressure equation, Eq.~\ref{eq:P}, the modification in the interactions between two bounding surfaces is wrought only through the changes in the ionic profiles and the electrostatic field. General forms of surface free energies may lead to surface phase separations as discussed in Ref.~\cite{podgornik2}.
As shown in the next section, such phase separation can couple to the spacing {\it between} the charged layers at equilibrium.

%%%%%%%%%%%%%%%%%%%%%%%%%%%%%%%%%%%%%%%%%%%%%%%%%%%%%%%%%%%%%%%%%%%%%%%%%%%
\subsection*{Lamellar-lamellar phase transition}
%%%%%%%%%%%%%%%%%%%%%%%%%%%%%%%%%%%%%%%%%%%%%%%%%%%%%%%%%%%%%%%%%%%%%%%%%%%
One extension of the Ninham-Parsegian model considers short-range non-electrostatic interactions
between associated and dissociated charged groups at the surface, and can be modeled
by additional terms to $f_s$~\cite{Daniel1}. This generalization is motivated by experiments by Zemb et al.~\cite{zemb98}, where a lamellar-lamellar phase transition was observed in certain bilayer-forming lipids and surfactants,
such as DDA (didodecyldimethylammonium halides), for a series of three homologous halides counterions: Cl$^-$, Br$^-$ and I$^-$. A discontinuous transition was found for the inter-lamellar spacing $D$ as a function of applied osmotic stress, but only when counterions such as bromide were used.
In contrast, for chloride there is no phase transition and the isotherm, $P(D)$, follows the PB prediction,
while for iodide the lamellar phase did not disperse at all in water.

There is a probable link between the surface activity of ions and the lamellar-lamellar phase transition characterized by the discontinuous jump (for certain counterions) in the inter-bilayer separation. To suggest a possible explanation for these experiments, a free-energy approach can be used, with a surface term $f_s$ of the form
\begin{equation}
f_s=\sigma \psi_s +f_{\rm ent}  -
\frac{k_BT}{a^2}\left[ \alpha_s \eta + \frac{1}{2} \chi_s \eta^2  \right]
\end{equation}

The modification in $f_s$, as compared with the previous section, is included in the last two terms in the squared brackets. A non-electrostatic ion binding to the surface is modeled  by the linear $\alpha_s\eta$ term, while the interaction between the bound and dissociated groups is represented by the quadratic  $\frac{1}{2}\chi_s\eta^2$ term.

Minimizing the complete free energy leads to the regular PB equation inside the bulk solution and to a boundary condition in the form of the Langmuir-Frumkin-Davis adsorption isotherm~\cite{Davies}. The solution of the PB equation with this boundary condition completely determines the counterion density profile $n_i(z)$ and electrostatic potential, and yields the osmotic pressure $P$ of the form Eq. \ref{P_PB} as a function of separation $D$.

The analysis of the solution shows that if the parameter $\chi_s$ is large enough (typically $\approx 10 k_BT$), an in-plane phase transition can occur, with a coexistence region in the phase diagram between ion-adsorbed  and ion-depleted  phases. This transition, in turn, is coupled to the bulk phase transition causing a jump in the inter-lamellar spacing $D$ as the osmotic pressure is changed.
Qualitatively, the behavior of this lamellar system, seen for different ions, can be understood in terms of the specific counterion interaction with the charged surfactant bilayer. The Cl$^-$ counterion  always dissociates from the bilayer-forming
DDA$^+$ surfactant, resulting in PB-like behavior, and a continuous $P(D)$ isotherm. For Br$^-$, the dissociation is partial, leading to a first-order transition and a coexistence between the two lamellar phases of two different $D$ spacings. Finally, for the I$^{-}$ counterion, the ion always stays associated with the DDA$^{+}$ surfactant and there is not sufficient repulsive interaction that
can stabilize the swelling of the stack under any osmotic pressure.

%%%%%%%%%%%%%%%%%%%%%%%%%%%%%%%%%%%%%%%%%%%%%%%%%%%%%%%%%%%%%%%%%%%%%%%%%%%
\subsection*{Bilayer curvature effects}
%%%%%%%%%%%%%%%%%%%%%%%%%%%%%%%%%%%%%%%%%%%%%%%%%%%%%%%%%%%%%%%%%%%%%%%%%%%
So far we have discussed rigid and flat surfaces. But often, interfaces are composed of lipids or surfactants that can be deformed at some free-energy cost. The adsorption process can then lead to a surface restructuring and deformation at equilibrium. For a multi-component layer, the compositional and elastic degrees of freedom are in general coupled, leading to a complex variational problem. Ion specificity is then manifested through charge regulation, through interactions between mobile associated and dissociated surface sites, and through other surface mediated interactions, such as bilayer elasticity. This is particulary relevant when considering the
adsorption of large macroions onto oppositely charged layers.

Phenomenological terms added to the free energy are useful here too. The free energy can be written as a sum $f_s=f_{\rm ent}+f_{\rm els}$. The additional term corresponds to an elastic energy in its Helfrich form, using the bending modulus $\kappa$ of the layer,
\begin{equation}
f_{\rm els}=\frac{1}{2} \kappa \left[ c-c_0(\eta) \right]^2
\end{equation}
where $c$ denotes the layer curvature, and $c_0(\eta)$ is the spontaneous curvature, depending on the local fraction of dissociated surfaces sites, $\eta$.

Using these modified equations, it has been possible to predict how charged macromolecules, such as DNA, proteins, and even viruses, can reshape
oppositely charged membranes, and sometimes induce a macroscopic phase separation from a flat lamellar phase to a deformed one~\cite{deformed}.

%%%%%%%%%%%%%%%%%%%%%%%%%%%%%%%%%%%%%%%%%%%%%%%%%%%%%%%%%%%%%%%%%%%%%%%%%%%
\subsection*{Surface tension of electrolyte solutions}
\label{sec:surface}
%%%%%%%%%%%%%%%%%%%%%%%%%%%%%%%%%%%%%%%%%%%%%%%%%%%%%%%%%%%%%%%%%%%%%%%%%%%
Another approach to study ion-surface interaction is to consider a non-electrostatic external field that has a direct dependence on the distance from the surface, and add this term ``by hand" to the free energy, ending up with modified ionic profiles. The main difference between this method and those mentioned in the previous sections is the assumption of a certain functional form of the non-electrostatic external field, which is independent of other system properties. One famous example to this method is the Onsager-Samaras theory for the surface tension of electrolytes~\cite{OS}. In this seminal work it was assumed that the image-charge potential can be regarded as an external potential that scales as $\psi_{\rm im}\sim\exp(-2\lambda_\mathrm{D}^{-1} z)/z$ where $z$ is the distance from the surface and $\lambda_\mathrm{D}$ is the Debye screening length. Taking into account $\psi_{\rm im}$ leads to a depletion layer in the vicinity of neutral interfaces.

In recent works~\cite{YLevin1,YLevin2,YLevin3}, a model was proposed to explain the adsorption to an air/water interface of ``soft" and large anions, such as iodide and bromide, and the depletion of ``hard" and small cations, such as sodium. The model was based on two additional and independent {\it external} fields, absent in the standard PB theory. The contribution of the first stems from the cost of the electrostatic energy of a polarizable ion that is modeled as a sphere, positioned at the interface and immersed only partially in the liquid phase. This energy is found to be substantially reduced  for a large soft anion. The second additional ingredient is the energy associated with exchanging an ion with a water molecule in the hydrogen-bond network. Since the water molecule is energetically favored in the bulk, this will lead to an effective attraction of the ion to the interface, which scales with the partial volume of the ion in the air phase.

Taking these two independent {\it external} fields into account leads to a dependence of the profiles on the ionic size and polarizability. The predictions suggest that the ion-specific surface tension agrees with the Hofmeister series~\cite{Kunz}.

In another work it was suggested that the specific ion property of making or breaking the water hydrogen-bond network can be modeled by an attractive potential (square well) or a repulsive potential (square barrier)~\cite{Ruckenstein1,Ruckenstein2}. Namely, the energetic favor or disfavor of ions to be part of the hydrogen-bond network leads to an effective attractive or repulsive interaction of the ions to the surface, respectively. Adding these external fields changes the ionic profile and induces an ion-specific depletion or adsorption, depending on the choice of the external potential. This non-monotonic behavior was also observed in surface tension experiments \cite{exp_surf_tension}.

It is still questionable whether the simplified form of these external potentials can be fully justified. Furthermore, in all models where independent external potentials are ``inserted by hand" into the free energy, it is assumed that the additional interactions are additive. It remains an open question whether all these interactions are not intimately related to one another in a self consistent manner.

%%%%%%%%%%%%%%%%%%%%%%%%%%%%%%%%%%%%%%%%%%%%%%%%%%%%%%%%%%%%%%%%%%%%%%%%%%%
\subsection*{Dispersion forces close to dielectric discontinuities}
%%%%%%%%%%%%%%%%%%%%%%%%%%%%%%%%%%%%%%%%%%%%%%%%%%%%%%%%%%%%%%%%%%%%%%%%%%%
Electrostatic interactions mediated by aqueous solvent are contingent upon the dielectric response function that encodes all the relevant structural properties of the solvent~\cite{dogonadze}. The temporal dependence is always non-local and can be converted into an appropriate frequency-dependent dielectric response function, $\varepsilon(\omega)$. The non-local spatial dependence, on the other hand, describes the orientational correlations between water molecules in a hydrogen-bonded network. For an infinite system it can be codified by a wave-vector dependent dielectric function~\cite{cevc}. However, additional approximations are needed in order to describe correlations in a system that is bound by interfaces and contains mobile counterions~\cite{horny}.

The link between the frequency-dependent $\varepsilon(\omega)$ and the van der Waals (vdW) interactions is provided by the Lifshitz theory~\cite{parsegianbook}. The dielectric models for the frequency response of aqueous solvents are well worked out and lead to quantitative estimates of these interactions for various geometries~\cite{RMP}. Ninham and Mahanty were the first to suggest that the vdW interactions supply, at least in part, an explanation to the ionic specificity~\cite{Ninhambook}. In particular, they suggested that the vdW interactions partially explain the origin of the water-structure contribution to the interactions between ions and surfaces. For ions interacting with surfaces the vdW potential in the non-retarded limit is given by $\psi_{\rm vdW}(z)\sim B/z^{3}$ where $z$ is the distance from the surface, and the coefficient $B$ depends on the polarizability of the ions. This suggestion has been picked up recently in several attempts to connect the Hofmeister series and other ion-specific effects with the vdW interactions (see Refs.~\cite{Kunz,Hoffmei} and references therein).

At the most basic level the vdW interactions contribute an additional term to the mean-field potential leading to modified ionic profiles compared to the PB ones~\cite{Edwards}. A more advanced mean-field continuum theory should take into account the dependence of the dielectric response on the local density of the ions~\cite{Dan2}. Thus, the frequency-dependent dielectric response could be written as $\varepsilon(z, \omega) = \varepsilon(n_i(z), \omega)$. Note that the ion density dependence of the dielectric function has already been discussed in the previous sections, but the evaluation of the vdW interactions with a non-homogeneous dielectric profile is, in general, difficult~\cite{Veble1}.

An additional complication is due to the fact that not only the vdW interactions depend on $n_i(z)$, but the converse is true as well --- the ionic profiles, $n_i(z)$, depend on vdW interactions. This coupled problem is even more difficult to solve and, in general, may lead to non-monotonic density profiles~\cite{Veble2}.

The vdW interactions add ionic specificity to the total free energy since $\varepsilon(n_i(z), \omega)$ depends crucially on the ion species. It is important, though, to note that this is only one part of the ion-specific effect. The other part has to be implemented within PB theory itself by consistently assuming that the static dielectric response depends on the ionic densities too, as advocated in previous sections of this review. Only taken {\it together}  these two effects can constitute a consistent formulation.

%%%%%%%%%%%%%%%%%%%%%%%%%%%%
\section{Concluding remarks}
%%%%%%%%%%%%%%%%%%%%%%%%%%%%
Ion-specific effects are present in a broad range of chemical and biological systems. These effects are, however, difficult to model since they usually involve complicated couplings that emerge from a large number of species and interactions. Here we have explored several models with a common theme --- all are phenomenological and highlight one or two aspects of ion-specific interactions. Furthermore, all the models are discussed within a common theoretical framework on the mean-field level, leading to relatively simplified and intuitive picture of ion-specific effects. The drawback of this approach is the lack of accuracy and microscopic description that can be provided by other methods, such as simulations and detailed liquid-state theories.

The various models can be grouped into two main types of non-electrostatic interactions:  bulk and surface ones. For the first type
we reviewed three non-electrostatic interactions.

{\it (i)~}The finite-size of ions, when non-negligible, leads to steric repulsion and to an upper bound on the ionic density that accumulates in the vicinity of charged surfaces. As a consequence, the width of the double-layer is increased for larger ions and the osmotic pressure in between two surfaces, separated by a few nanometers, grows.

{\it (ii)}~Ions influence the dielectric response in their vicinity. Mostly, ions apply strong electrostatic field on neighboring water molecules, leading to a local decrement of the dielectric response. This, in turn, leads to an additional effective interaction of ions with charged surfaces. The effect on the ion density profile is somewhat similar to that of large ions in the steric interaction model, and the double layer width is increased. Ion specificity in this model emerges via the strength of the ionic influence on the local dielectric response.

{\it (iii)~}When the solvent is composed of several components, additional interactions lead to modifications of the ionic and solvent density profiles. In the model described above, the change of local dielectric response due to local changes of solvent composition, and the difference in the solvation energy of the ions in different solvents lead to a decrease in the osmotic pressure as compared to a homogeneous solvent composition.

The second type of non-electrostatic interactions are surface interactions and several cases were discussed.

{\it (i)~}Surface charge can be self-adjusted by various mechanisms, such as ionization and ionic dissociation. Such self-adjusted surfaces interact with the ionic solution in a self-consistent manner, which leads to changes in the ionic density profiles and osmotic pressure.

{\it (ii)~}Adding an in-plane non-electrostatic interaction of the surface species to the charge regulation mechanism may further lead to a phase transition as a function of the osmotic pressure between two charged surfaces. Furthermore, when coupling the charge regulation mechanism to the membrane curvature degrees of freedoms, structural changes in the membrane can be observed when it is brought in contact with an ionic solution.

{\it (iii)~}Considering long-range external surface interactions such as dispersion interaction, that result from the dielectric inhomogeneities in the system, leads to further modifications in ionic density profiles and osmotic pressure. Here too, the ionic specificity emerges from the ion-dependent dielectric response of the solution.

Resolving a complete picture of ion-specific effects remains an ever-challenging task. In order to accomplish this task, it will be
helpful  to obtain first a better understanding of each of the myriad experimental observations in terms of models based on simple ingredients. Then, combining the separate pieces of knowledge gained together into a single and complete picture of ion-specific interactions may hopefully be achieved.

{\em Acknowledgement:~~~} Partial support from the Slovenian Research Agency through program P1-0055 (research project J1-0908), the Isreali-Slovene joint research project,  the Israel Science Foundation (ISF) under grant no. 231/08, and the US--Israel
Binational Science Foundation (BSF) under grant no. 2006/055 is gratefully acknowledged. RP would like to thank 
the Aspen Center for Physics where parts of this work have been conceived
during the workshop {\sl New Perspectives in Strongly Correlated Electrostatics in Soft Matter}.

%%%%%%%%%%%%%%%%%%%%%%%%%%

%%%%%%%%%%%%%%%%%%%%%

%%%%%%%%%%%%%%

\begin{thebibliography}{9}
\bibitem{Thefrenchguy}
We are indebted to MZ Bazant for providing us with historical details regarding this issue.
\end{thebibliography}
\end{document}